\documentclass[12pt,preprint]{aastex}
\usepackage{amsmath}

\begin{document}

\title{GeV Emission during X-Ray Flares from Late Internal Shocks: Application to GRB 100728A}

\author{K. Wang$^{1,2}$ and Z. G. Dai$^{1,2}$}
\affil{$^{1}$School of Astronomy and Space Science, Nanjing University, Nanjing 210093, China; dzg@nju.edu.cn \\
$^{2}$Key Laboratory of Modern Astronomy and Astrophysics (Nanjing
University), Ministry of Education, China \\}

\begin{abstract}
Recently, the GeV radiation during the X-ray flare activity in GRB
100728A was detected by Feimi/LAT. Here we study the dynamics and
emission properties of collision between two homogeneous shells
based on the late internal shock model. The GeV photons can be
produced from X-ray flare photons up-scattered by relativistic
electrons accelerated by forward-reverse shocks, where involved
radiative processes include synchrotron self-Compton and crossing
inverse-Compton scattering. By analytical and numerical
calculations, the observed spectral properties in GRB 100728A can be
well explained.
\end{abstract}

\keywords{gamma rays: bursts --- radiation mechanisms: non-thermal}

\section{Introduction}
Gamma-ray bursts (GRBs) are the brightest explosive phenomena in the
universe, the study of which has been being one of the most
interesting fields in astrophysics. Thanks to the launch of the
Fermi satellite in 2008, the Large Area Telescope (LAT) onboard
Fermi has detected high-energy photons in energy range from 20 MeV
to 300 GeV. Several mechanisms have been proposed to predict the
origin of GeV photons along with the GRB afterglow phase (for a
review see Zhang 2007): (1) In the external shock scenario,
high-energy photons may be produced by synchrotron radiation and
synchrotron self-Compton (SSC) processes from forward-reverse shocks
\citep{mes94a, mes94b, der00, zhang01, sar01} or crossing
inverse-Compton (CIC) processes between forward-reverse shocks
\citep{wang01a, wang01b, peer05}. (2) In the hadronic and photo-pion
scenario, there may be synchrotron radiation of protons, $\pi^+$
from $p\gamma$, $pn$ and $pp$ interactions, and positrons produced
from $\pi^+$ decay and $\pi^0$ decay from $p\gamma$ interactions
\citep{gupta07}. (3) Electrons from pair productions during
interaction of $>$100 GeV photons from GRBs with cosmic infrared
background photons might also emit GeV photons by inverse scattering
off cosmic microwave background photons \citep{dai02, wang04}.

On the other hand, one of the key discoveries is bright X-ray flares
superimposing on underlying afterglow emission from nearly a half of
GRBs observed by Swift \citep{bur05}. The rapid rise and decay
behavior of X-ray flares is widely understood as being due to some
long-lasting activity of the central engines. Such an activity might
be caused by an instable accretion disk around a black hole
\citep{perna06}, accretion of fragments of the collapsing stellar
core onto a central compact object in the collapsar model
\citep{king05}, a modulation of accretion flow by a magnetic barrier
\citep{proga06}, or magnetic reconnection of a newborn neutron star
\citep{dai06}.

GRB 100728A is a case with simultaneous detections by Swift and
Fermi \citep{abd11}, which detected GeV photons during X-ray flares.
The GeV photons during the X-ray flare activity detected by
Fermi/LAT have been thought to arise from external inverse-Compton
(EIC) scattering off X-ray flare photons by electrons in a
relativistic forward shock \citep{fan06, fan08, wang06, he12}. Here
we propose a different explanation, in which the detected GeV
photons are produced by SSC and CIC scattering off X-ray flare
photons by electrons accelerated in the late internal shock model.
This model was suggested by \citet{fan05} and \citet{zhang06}, and
its motivations are based on two following facts. First, the rapid
rising and decaying timescales and their distributions of X-ray
flares require that the central engine restarts at a later time
\citep{lazzati07}. Second, \citet{liang06} fitted the light curves
of X-ray flares detected by Swift by assuming that the decaying
phase of an X-ray flare is due to the high latitude emission from a
relativistic outflow. These authors found that the ejection time of
this outflow from the central engine is nearly equal to the peak
time of an observed X-ray flare produced by the outflow.

This paper is organized as follows: we calculate the dynamics of a
collision between two shells and properties of synchrotron and IC
emission to produce X-ray flares and higher-energy emissions in
section 2. In section 3, we present numerical calculations and light
curves of the model. In this section, we also make an application to
GRB 100728A and present constraints on the model parameters.  In the
final section, some conclusions are given.

\section{The Synchrotron and IC Emission from Late Internal Shocks}

In the internal shock model, a fireball consisting of a series of
shells with different Lorentz factors can form prompt emission
through shell-shell interactions. Similarly, collisions between
shells with different velocities ejected from the central engine at
late times after the GRB trigger can form late internal shocks, the
emission from which reproduce X-ray flares.

\subsection{Dynamics of Two Shell Collisions}
For one X-ray flare we here consider the following shell-shell
collision: a prior slow shell A with bulk lorentz factor
$\gamma_{A}$ and kinetic-energy luminosity $L_{k,A}$, and a
posterior fast shell B with bulk lorentz factor $\gamma_{B}$
$(\gamma_{A}<\gamma_{B})$ and kinetic-energy luminosity $L_{k,B}$.
The collision of the two shells takes place at radius
\begin{equation}
{R_{col}} = {\beta _B}c\frac{{{\beta _A}\Delta {t_{ej}}}}{{({\beta _B} -
{\beta _A})}} \simeq \frac{{2\gamma _A^2c\Delta {t_{ej}}}}{{1 -
{{({\gamma _A}/{\gamma _B})}^2}}} \equiv 2\gamma _A^2c\delta t,
\end{equation}
where $\Delta {t_{ej}}$ is the ejection interval of the two shells,
and $\delta t$ is redefined interval. During the collision, there
are four regions separated by forward-reverse shocks: (1) the
unshocked shell A, (2) the shocked shell A, (3) the shocked shell B,
and (4) the shocked shell B, where regions 2 and 3 are separated by
a contact discontinuity.

The particle number density of a shell measured in its comoving
frame can be calculated by
\begin{equation}
{n'_i} = \frac{{{L_{k,i}}}}{{4\pi {R^2}\gamma _i^2{m_p}{c^3}}},
\end{equation}
where $R$ is the radius of the shell and subscript $i$ can be taken
as A or B.

\citet{yu09} had analyzed the dynamics of a late-time shell-shell
collision in detail. In order to get a high theoretical X-ray
luminosity, it is reasonable to assume $\gamma_{A}\ll\gamma_{B}$ and
$L_{k,A}=L_{k,B}\equiv L_{k}$. Assume that $\gamma_{1}$,
$\gamma_{2}$, $\gamma_{3}$, and $\gamma_{4}$ are Lorentz factors of
regions 1, 2, 3 and 4, respectively. As a result, we have
$\gamma_{1}=\gamma_{A}$, $\gamma_{4}=\gamma_{B}$, and $n'_1\gg
n'_4$. If a fast shell with low particle
number density catches up with a slow shell with high particle
number density and then collides with each other, a Newtonian
forward shock (NFS) and a relativistic reverse shock (RRS) may be
generated \citep{yu09}. So we can obtain
$\gamma_{1}\simeq\gamma_{2}=\gamma_{3}=\gamma\ll\gamma_{4}$. Then,
according to the jump conditions between the two sides of a shock
\citep{bla76}, the comoving internal energy densities of the two
shocked regions can be calculated by ${e'_2} = ({\gamma _{21}} -
1)(4{\gamma _{21}} + 3){n'_1}{m_p}{c^2},{e'_3} = ({\gamma _{34}} -
1)(4{\gamma _{34}} + 3){n'_4}{m_p}{c^2}$, where
$\gamma_{21}=\frac{1}{2}({\gamma _1}/{\gamma _2} + {\gamma
_2}/{\gamma _1})$ and $\gamma_{34}=\frac{1}{2}({\gamma _3}/{\gamma
_4} + {\gamma _4}/{\gamma _3})$ are the Lorentz factors of region 2
relative to the unshocked region 1, and region 3 relative to region
4, respectively. It is required that $e'_2=e'_3$ because of the
mechanical equilibrium. We have
\begin{equation}
\frac{{({\gamma _{21}} - 1)(4{\gamma _{21}} + 3)}}{{({\gamma _{34}} - 1)
(4{\gamma _{34}} + 3)}} = \frac{{n'_4}}{{n'_1}}
={\left( {\frac{{{\gamma _1}}}{{{\gamma _4}}}} \right)^2} \equiv f.
\end{equation}
Two relative Lorentz factors can be calculated by ${\gamma _{21}}
\approx \frac{{f\gamma _4^2}}{{7\gamma _1^2}} + 1 = \frac{8}{7}$,
and ${\gamma _{34}} = \frac{{{\gamma _4}}}{{2{\gamma _1}}} \gg 1$.
Assuming that $t$ is the observed shell interaction time since the
X-ray flare onset, the radius of the system after the collision can
be written as
\begin{equation}
R = {R_{col}} + 2{\gamma ^2}ct \simeq 2{\gamma ^2_1}c(t + \delta t).
\label{R}
\end{equation}

During the propagation of the shocks and before the shock crossing,
the total electron numbers in regions 2 and 3 can be calculated by
${N_{e,2}} = 8\pi {R^2}{n'_1}({\gamma _{21}}{\beta _{21}}/\gamma
\beta ){\gamma ^2}ct$ and ${N_{e,3}} = 8\pi {R^2}{n'_4}({\gamma
_{34}}{\beta _{34}}/\gamma \beta ){\gamma ^2}ct$ \citep{dai02},
respectively. We can easily find that the electron number in region
2 is larger than that in region 3.

\subsection{Synchrotron Emission from Two Shocked Regions}

As usual, we assume that fractions of $\epsilon_{B}$ and
$\epsilon_{e}$ of the internal energy density in a GRB shock are
converted into the energy densities of the magnetic field and
electrons, respectively. Thus, using ${B'_i} = {(8\pi {\epsilon
_B}{e'_i})^{1/2}}$ for $i=$2 or 3, the strength of the magnetic
field is calculated by
\begin{equation}
{B'_2} ={B'_3} = {\left[\frac{{{\epsilon_B}L_k}}{{2{\gamma
^6_1}{c^3}{{(t + \delta t)}^2}}}\right]^{1/2}}.
\end{equation}

The electrons accelerated by the shocks are assumed to have a
power-law energy distribution, $d{n'_{e,i}}/d{\gamma '_{e,i}}
\propto {\gamma '_e}^{ - p}$ for ${\gamma '_{e,i}} \ge {\gamma
'_{e,m,i}}$, where ${\gamma '_{e,m,i}}$ is the minimum Lorentz
factor. According to ${\gamma '_{e,m,i}} =
\frac{{{m_p}}}{{{m_e}}}(\frac{{p - 2}}{{p - 1}}){\epsilon
_e}({\gamma _{\rm rel}} - 1)$ (where ${\gamma _{\rm
rel}}=\gamma_{21}$ or $\gamma_{34}$ in region 2 or 3), the minimum
Lorentz factor can be written as
\begin{equation}
\begin{array}{c}
{\gamma '_{e,m,3}}
 \simeq 2.8 \times {10^3}{g_{p}}{\epsilon _{e, - 1/2}}{\gamma _{4,2.5}}\gamma _{1,1}^{ - 1},
\end{array}
\end{equation}
\begin{equation}
\begin{array}{c}
{\gamma '_{e,m,2}}
 \simeq 30{g_{p}}{\epsilon _{e, - 1/2}},
\end{array}
\end{equation}
where $\epsilon _{e, - 1/2}=\epsilon _{e}/10^{-1/2}$, $\gamma
_{4,2.5}=\gamma _{4}/10^{2.5}$, $\gamma _{1,1}=\gamma _{1}/10^1$, and
$g_p=3(p-2)/(p-1)$.

Moreover, the cooling Lorentz factor, above which the electrons lose
most of their energies, ${\gamma '_{e,c,i}} = 6\pi
{m_e}c/({y_i}{\sigma _T}{B'_3}^2\gamma t)$, should be given by
\begin{equation}
{\gamma '_{e,c,3}} = {\gamma '_{e,c,2}} \simeq 1.4 \times {10^3}
y_{,0}^{ - 1}\varepsilon _{B, - 3/2}^{ - 1}L_{k,50}^{ - 1}\gamma _{1,1}^5
\frac{{(t + \delta t)_{,2}^2}}{{{t_{, - 2}}}}.
\end{equation}
where $y_i=1+Y_i$ is the ratio of the total luminosity to
synchrotron luminosity, and
$Y_i\approx[(4\eta_{i}\epsilon_e/\epsilon_B+1)^{1/2}-1]/2$ is the
Compton parameter, which is defined by the ratio of the IC to
synchrotron luminosity, with $\eta_{i}=\min[1,({\gamma
'_{e,c,i}}/{\gamma '_{e,m,i}})^{2-p}]$ \citep{sar01}. Here we assume
$\epsilon_e=0.3$ and $\epsilon_B=0.03$ in our calculations, so
$y_i<3$ can be easily obtained so that we can assume $y_i\sim 1$.
Fig. 1 presents changes of $Y_i$ and shows that it is reasonable to
assume $y_2\sim y_3\sim 1$. Thus, the IC luminosity is comparable
with the synchrotron luminosity.

In order to obtain the synchrotron emission spectrum, we consider
\begin{equation}
{\nu _{m,i}} = \frac{{{q_e}}}{{2\pi {m_e}c}}{B'_i}{\gamma '} ^2_{e,m,i}\gamma ,
\end{equation}
and
\begin{equation}
{\nu _{c,i}} = \frac{{{q_e}}}{{2\pi {m_e}c}}{B'_i}{\gamma '} ^2_{e,c,i}\gamma ,
\end{equation}
where $q_{e}$ is the electron charge. Four characteristic
frequencies in regions 2 and 3,
\begin{equation}
{\nu _{m,2}} \simeq 4.5 \times {10^{13}} g_{p}^2\epsilon _{e, -
1/2}^2\epsilon _{B, - 3/2}^{1/2}L_{k,50}^{1/2}\gamma _{1,1}^{ - 2}(t
+ \delta t)_{,2}^{ - 1} \; \mathrm{Hz},\label{eqnum2}
\end{equation}
\begin{equation}
{\nu _{m,3}} \simeq 5.0 \times {10^{17}} g_{p}^2\epsilon _{e, -
1/2}^2\epsilon _{B, - 3/2}^{1/2}L_{k,50}^{1/2}\gamma
_{4,2.5}^2\gamma _{1,1}^{ - 4}(t + \delta t)_{,2}^{ - 1} \; \mathrm{Hz},\label{eqnum3}
\end{equation}
and
\begin{equation}
{\nu _{c,2}}={\nu _{c,3}} \simeq 1.3 \times {10^{17}}y_{,0}^{ -
2}\epsilon _{B, - 3/2}^{ - 3/2}L_{k,50}^{ - 3/2}\gamma
_{1,1}^8\frac{{(t + \delta t)_{,2}^3}}{{t_{, - 2}^2}} \; \mathrm{Hz},\label{eqnuc}
\end{equation}
can be obtained. In Fig. \ref{f2}, their time evolutions are
presented. From this figure, we can know easily that region 2 and
region 3 are in the slow cooling regime at very early times,
subsequently region 2 in the slow cooling regime but region 3 in the
fast cooling regime, and finally both regions in the fast cooling
regime. As a result, the spectral index between $\nu_{m}$ and
$\nu_{c}$ of region 2 and region 3 has an evolution with time as
\citet{sar98}. It is reasonable that region 3 can be thought to be
in the fast cooling regime, while region 2 is in the slow cooling
regime at early times and in the fast cooling regime at later times.
By applying the formula
\begin{equation}
{F_{\nu ,\max ,i}} = \frac{{{N_{e,i}}}}{{4\pi D_L^2}}\frac{{{m_e}{c^2}{\sigma _T}}}{{3{q_e}}}{B'_i}\gamma ,
\end{equation}
where $D_{L}$ is the luminosity distance of the burst, we obtain the
peak flux density
\begin{equation}
{F_{\nu ,\max ,2}} \simeq 0.11 \epsilon_{B, - 3/2}^{1/2}
L_{k,50}^{3/2}\gamma _{1,1}^{ - 3}\frac{{{t_{, - 2}}}}{{{{(t + \delta t)}_{,2}}}}
D_{L,28}^{ - 2} \; \mathrm{Jy},   \label{fnumax2}
\end{equation}
and
\begin{equation}
{F_{\nu ,\max ,3}} \simeq 1.6 \times {10^{ - 3}} \epsilon _{B, -
3/2}^{1/2}L_{k,50}^{3/2}\gamma _{4,2.5}^{ - 1}\gamma _{1,1}^{ -
2}\frac{{{t_{, - 2}}}}{{{{(t + \delta t)}_{,2}}}}D_{L,28}^{ - 2} \; \mathrm{Jy}.  \label{fnumax3}
\end{equation}
According to equations (\ref{fastcooling}) and (\ref{slowcooling})
in appendix A \citep{sar98}, the synchrotron spectrum of region 2 in
the slow cooling regime ($\nu_{m,2}<\nu_{c,2}$) is thus described by
${F_{\nu ,2}} = {F_{\nu ,\max ,2}}{(\nu /{\nu _{m,2}})^{ - (p -
1)/2}}$ for ${\nu _{m,2}} < \nu  < {\nu _{c,2}}$ and ${F_{\nu ,2}} =
{F_{\nu ,\max ,2}} {({\nu _{c,2}}/{\nu _{m,2}})^{ - (p - 1)/2}}{(\nu
/{\nu _{c,2}})^{ - p/2}}$ for $\nu  > {\nu _{c,2}}$ or in the fast
cooling regime ($\nu_{c,2}<\nu_{m,2}$) by ${F_{\nu ,2}} = {F_{\nu
,\max ,2}}{(\nu /{\nu _{c,2}})^{ -1/2}}$ for ${\nu _{c,2}} < \nu  <
{\nu _{m,2}}$ and ${F_{\nu ,2}} = {F_{\nu ,\max ,2}} {({\nu
_{m,2}}/{\nu _{c,2}})^{ - 1/2}}{(\nu /{\nu _{m,2}})^{ - p/2}}$ for
$\nu  > {\nu _{m,2}}$. In the fast cooling regime of region 3,
${F_{\nu ,3}} = {F_{\nu ,\max ,3}}{(\nu /{\nu _{c,3}})^{ - 1/2}}$
for ${\nu _{c,3}} < \nu < {\nu _{m,3}}$ and ${F_{\nu ,3}} = {F_{\nu
,\max ,3}}{({\nu _{m,3}} /{\nu _{c,3}})^{ - 1/2}}{(\nu /{\nu
_{m,3}})^{ - p/2}}$ for $\nu  > {\nu _{m,3}}$.

\subsection{IC Emission from Two Shocked Regions}
\begin{figure}
\epsscale{.96} \plotone{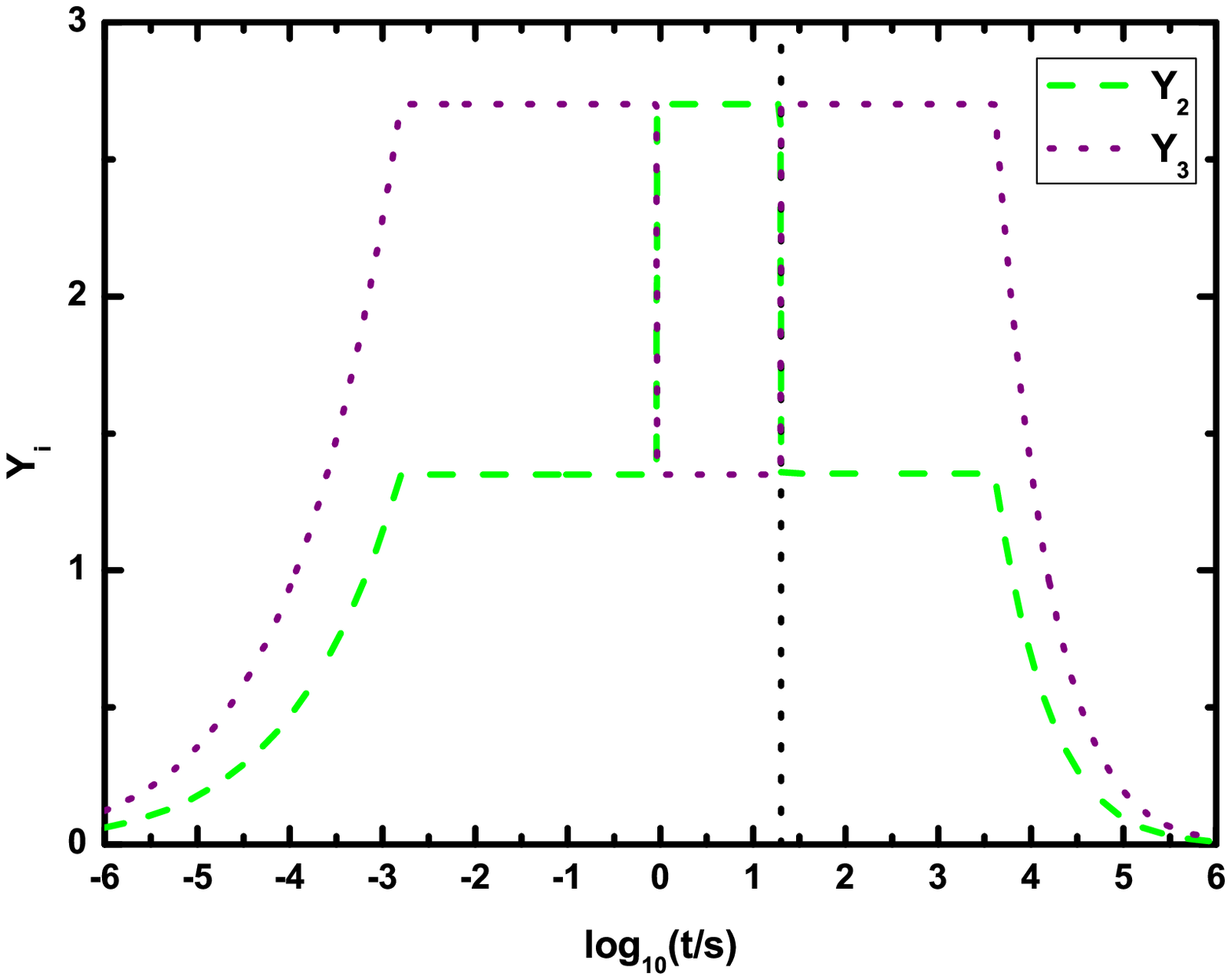} \caption{Ratio $Y_{i}$ of the IC
to synchrotron luminosity as a function of time. The black dotted
line represents the forward and reverse shock crossing time. Here we
assume that the two shocks cross the two shells at a similar time
$T_{crs}=20\,\mathrm{s}$. After the shock crossing time, the merged
shell expands adiabatically if $s=3$ is assumed. The other
parameters $L_{k,1}=L_{k,4}=10^{50}\mathrm{erg s^{-1}}$,
$\gamma_{1}=10$, $\gamma_{4}=300$, $p=2.5$, $\epsilon_e=0.3$,
$\epsilon_B=0.03$, $\theta_{jet}=0.1$, and $z=1$ are taken in
numerical calculations.\label{f1}}
\end{figure}

The ratio of IC to synchrotron emission luminosity $Y_{i}$ has been
mentioned above (Fig. \ref{f1}). Although regions 2 and 3 forming
during the two-shell collision are optically thin to electron
scattering, some synchrotron photons will be Compton scattered by
shock-accelerated electrons, producing an additional IC component at
higher-energy bands. Considering the highest energy electrons whose
scattering enters the Klein-Nishina (KN) regime, the KN break
frequency is calculated by
\begin{equation}
h\nu _{KN,3}^{SSC} = \frac{{{\gamma ^2}m_e^2{c^4}}}{{h{\nu _{m,3}}}}
\simeq 13 g_{p}^{-2}\epsilon _{e, -1/2}^{-2}\epsilon _{B, -
3/2}^{-1/2}L_{k,50}^{-1/2}\gamma _{4,2.5}^{  -2}\gamma _{1,1}^{  6}(t
+ \delta t)_{,2} \; \mathrm{GeV}.\label{kn}
\end{equation}

Because of the characteristic frequency $h\nu_{m,3}\sim1$\,keV, and
$\gamma '_{e,m,3}\sim10^3$, we can obtain $(\gamma'_{e,m,3} h
\nu_{m,3})/(m_e c^2) \sim 1 $. So in the analysis estimates, it is
reasonable to use the Thomson optical depth of the electrons in
regions 2 and 3, which can be calculated by ${\tau _i} =
\frac{{{\sigma _T}{N_{e,i}}}}{{4\pi {R^2}}}$, where $i=2$ or $3$. We
calculate the upscattered spectral characteristic frequencies of IC
process, as in \citet{sar01}. Region 3 is in the fast cooling regime
and its SSC break frequencies become
\begin{equation}
  h\nu _{m,3}^{ssc} = 2{\gamma '} ^2_{e,m,3} h{\nu _{m,3}}
    \simeq 32 g_{p}^4\epsilon _{e, - 1/2}^4\epsilon _{B, - 3/2}^{1/2}
     L_{k,50}^{1/2}\gamma _{4,2.5}^4\gamma _{1,1}^{ - 6}(t + \delta t)_{,2}^{ - 1} \; \mathrm{GeV},\label{eqnum3ssc}
\end{equation}
and
\begin{equation}
  h\nu _{c,3}^{ssc} = 2{\gamma '} ^2_{e,c,3} h{\nu _{c,3}}
  \simeq 2.1 y_{,0}^{ - 4}\epsilon _{B, - 3/2}^{ - 7/2}
   L_{k,50}^{ - 7/2}\gamma _{1,1}^{18}\frac{{(t + \delta t)_{,2}^7}}{{t_{, - 2}^4}} \; \mathrm{GeV}.
\end{equation}

Obviously, the SSC peak energy for region 3 is in the KN regime and
$h\nu _{m,3}^{ssc}$ is comparable with $h\nu _{KN,3}^{SSC}$. As
\citet{tav98} suggested, no matter whether the SSC peak frequency
enters the KN regime or not, the spectral index of SSC emission at
low energy band has the same power-law approximation as synchrotron
emission. So the SSC flux of the fast-cooling region 3, ${F_{\nu
,3}^{SSC}} = {F_{\nu ,\max ,3}^{SSC}}{(\nu /{\nu _{c,3}^{SSC}})^{ -
1/2}}$ for ${\nu _{c,3}^{SSC}} < \nu  < \nu_{cri}$, where $\nu_{cri}
\sim \min({\nu_{m,3}^{SSC}}$, ${\nu_{KN,3}^{SSC}}$). As a result,
the peak flux at $\nu _{KN,3}^{SSC}$ is
\begin{eqnarray}
[\nu {F_\nu }]_{p,3}^{SSC} &=& \nu _{KN,3}^{SSC}{\tau _3}{F_{\nu ,\max ,3}}{\left(\frac{{\nu _{KN,3}^{SSC}}}{{\nu _{c,3}^{ssc}}}\right)^{ - 1/2}}     \nonumber \\
&\simeq& 6.7 \times {10^{ - 9}} g_p^{-1} y_{,0}^{ - 2}
 \epsilon _{e, - 1/2}^{ - 1} \epsilon _{B, - 3/2}^{ - 3/2} L_{k,50}^{1/2}\gamma _{4,2.5}^{
- 3}\gamma _{1,1}^{6}  (t + \delta t)_{,2}  D^{-2}_{L,28} \; \mathrm{erg \, cm^{-2} s^{-1}} .\label{knflux}
\end{eqnarray}

Region 2 is in the slow cooling regime, its SSC break frequencies
are $\nu _{m,2}^{SSC} = 2{\gamma '} ^2_{e,m,2} {\nu _{m,2}} \simeq
8.1 \times {10^{16}}g_p^4\epsilon _{e, - 1/2}^4\epsilon _{B, -
3/2}^{1/2}L_{k,50}^{1/2} \gamma _{1,1}^{ - 2}{(t + \delta t)^{ - 1}}
\; \mathrm{Hz}$ and $\nu _{c,2}^{ssc} = \nu _{c,3}^{ssc}$. Thus we
can obtain a very low peak flux
\begin{eqnarray}
[\nu {F_\nu }]_{p,2}^{SSC} &=& \nu _{c,2}^{SSC}{\tau _2}{F_{\nu ,\max ,2}}
{\left(\frac{{\nu _{c,2}^{SSC}}}{{\nu _{m,2}^{SSC}}}\right)^{ - (p - 1)/2}}  \nonumber \\
&\simeq& 9.8 \times {10^{ - 11}}y_{,0}^{ -
1}g_{p}^3\epsilon _{e, - 1/2}^3L_{k,50}^2\gamma _{1,1}^{ - 5}\frac{{t_{, - 2}}}{{(t +
\delta t)_{,2}^2}}D^{-2}_{L,28} \; \mathrm{erg \, cm^{-2}s^{-1}} , \label{fnup2ssc}
\end{eqnarray}
where $p=2.5$ is assumed. Obviously, the SSC radiation of region 2
is much weaker than that of region 3.

Apart from the SSC scattering processes in regions 2 and 3, the two
other cross-IC scattering processes are also presented. Assuming the
thin shell approximation, about one-half of the photons produced in
one shocked region will diffuse into the other one in the comoving
frame. We can obtain the low and high characteristic frequencies in
the following cases.
\newline(1) The synchrotron photons in region 2 are scattered by electrons in region
3,
\begin{eqnarray}
\nu _{L,3}^{CIC} &=& 2{\gamma '} ^2_{e,c,3}{\nu _{m,2}} \nonumber\\
&\simeq& 1.76 \times {10^{20}}y_{,0}^{ - 2}g_{p}^2
\epsilon _{e, - 1/2}^2\epsilon _{B. - 3/2}^{ - 3/2}{L^{ - 3/2}}
 \gamma _{1,1}^8\frac{{(t + \delta t)_{,2}^3}}{{t_{, -
2}^2}} \; \mathrm{Hz},
\end{eqnarray}
\begin{eqnarray}
\nu _{H,3}^{CIC} &=& 2{\gamma '} ^2_{e,m,3}{\nu _{c,2}} \nonumber\\
&\simeq& 1.97 \times {10^{24}} y_{,0}^{ - 2}g_{p}^2
\epsilon _{e, - 1/2}^2\epsilon _{B. - 3/2}^{ - 3/2}{L^{ - 3/2}}
 \gamma _{4,2.5}^2\gamma _{1,1}^6\frac{{(t + \delta
t)_{,2}^3}}{{t_{, - 2}^2}} \; \mathrm{Hz},
\end{eqnarray}
and the peak flux at $\nu _{H,3}^{CIC}$ can be estimated to be $[\nu
{F_\nu }]_{p,3}^{CIC} \sim 1 \times {10^{ -
9}}\mathrm{erg \, cm^{-2}s^{-1}}$.
\newline(2) The synchrotron photons in region 3 are scattered by electrons in region 2,
\begin{eqnarray}
\nu _{L,2}^{CIC} &=& 2{\gamma '} ^2_{e,m,2} {\nu _{c,3}} \nonumber\\
&\simeq& 2.34 \times {10^{20}}y_{,0}^{ - 2}g_{p}^2
\epsilon _{e, - 1/2}^2\epsilon _{B. - 3/2}^{ - 3/2}{L^{ - 3/2}}
 \gamma _{1,1}^8\frac{{(t + \delta t)_{,2}^3}}{{t_{, -
2}^2}} \; \mathrm{Hz},
  \end{eqnarray}
\begin{eqnarray}
  \nu _{H,2}^{CIC} &=& 2{\gamma '} ^2_{e,c,2} {\nu _{m,3}} \nonumber\\
  &\simeq& 1.96 \times {10^{24}}y_{,0}^{ - 2}g_{p}^2
  \epsilon _{e, - 1/2}^2\epsilon _{B. - 3/2}^{ - 3/2}{L^{ - 3/2}}
   \gamma _{4,2.5}^2\gamma _{1,1}^6\frac{{(t + \delta t)_{,2}^3}}{{t_{, - 2}^2}} \; \mathrm{Hz} ,
\end{eqnarray}
and the peak flux at $\nu _{H,2}^{CIC}$ can be estimated to be $[\nu
{F_\nu }]_{p,2}^{CIC} \sim 1 \times {10^{ -
9}}\mathrm{erg \, cm^{-2}s^{-1}}$.

From the above equations and Fig. \ref{f3}, for synchrotron
emission, region 3 is more important than region 2. For IC emission,
we can also see that the SSC emission of region 3 is the most
important among the IC components, while the SSC emission of region
2 is the weakest. This is very easy to understand, since the
electrons in region 3 have larger Lorentz factors due to RRS but the
electrons in region 2 have smaller Lorentz factors due to NFS.

\section{Application to GRB 100728A and Numerical Calculations}

\subsection{Parameter Limits}
The Fermi/GBM triggered GRB 100728A at 02:17:31 UT, 53.6\,s before
the Swift/BAT trigger. The duration of this burst is
$T_{90}\sim163$\,s. Several apparent X-ray flares were observed by
Swift/XRT, while significant GeV photons were detected by Fermi/LAT
during the early afterglow phase. We can obtain the observed
properties of this GRB: 1) From XRT, the time-averaged spectrum of
these flares from $t\sim167$ s to $ 854$ s can be fitted well by the
Band function \citep{ban93} with the low energy slope of
$\alpha=-1.06\pm0.11$, the high energy slope $\beta=-2.24\pm0.02$
and the peak energy $E_{pk}=1.0_{-0.4}^{+0.8}\,\mathrm{keV}$
\citep{abd11}. 2) From LAT, the spectrum of the GeV emission is
fitted well with photon index of
$\Gamma_{LAT}=-1.4\pm0.2\,\,(1\sigma)$ \citep{abd11} and the flux
$F_{LAT}\sim(5.8\pm4.5)\times10^{-9}\,\mathrm{erg \, cm^{-2}s^{-1}}$
\citep{he12} during the period of $t\sim167$ s to $ 854$ s. We use
our model with reasonable parameters to fit the GRB100728A
time-averaged energy spectrum (Fig. \ref{f4}). The data points in
this figure are taken from \citet{abd11}, from $T_0
+254\,\mathrm{s}$ to $T_0 +854\,\mathrm{s}$ about $7$ flares, where
$T_0$ is the trigger time. The duration of one flare is about tens
of seconds. Taking into account the similarity among the flares
generated, we only model one flare induced by a collision between
two shells to fit the interval data, so we choose the time from the
onset of two-shell interaction, i.e., $t=0\,\mathrm{s}$ to
$t=10^{1.8}\,\mathrm{s}$, where the latter time is comparable with
the duration of one flare of GRB 100728A.

The emission of region 3 is the most important, and is used to
explain the observations on GRB100728A. Since region 3 is in the
fast cooling regime and the high energy slope $\beta=-2.24\pm0.02$,
we can obtain the electron distribution index $p=2.48\pm0.04$. For
${\nu _{c,3}} < \nu  < {\nu _{m,3}}$ and ${\nu _{c,3}^{SSC}} < \nu <
{\nu _{cri}}$, the synchrotron spectrum and SSC component of an
X-ray flare have the same photon index of $-3/2$, which is
consistent with the observed GeV emission,
$\Gamma_{LAT}=-1.4\pm0.2$. The low energy slope of
$\alpha=-1.06\pm0.11$, which may be caused by the low frequency
absorption effect, can also be regarded as a consistent result
within the acceptable range.

In the two-shell collision model, we only regard the kinetic-energy
luminosity $L_{k}$, Lorentz factors $\gamma_{1}$ and $\gamma_{4}$ as
variable parameters. Because of $h\nu_{m,3}\sim
E_{pk}=1.0_{-0.4}^{+0.8}\,\mathrm{keV}$ and $h\nu_{m,3}^{SSC}
\gtrsim h\nu_{KN,3}^{SSC} \sim h\nu_{pk}^{LAT}>10\,$GeV, from the
ratio of equations (\ref{eqnum3}) and (\ref{eqnum3ssc}),
$\gamma_{4}/\gamma_{1}>30$ is required, which is consistent with the
dynamical analysis. This suggests that the posterior shell can catch
up with the prior shell very soon and an NFS and a RRS can be
formed. Furthermore, according to equation (\ref{knflux}) and
$F_{LAT}\sim(5.8\pm4.5)\times10^{-9}\,\mathrm{erg \,
cm^{-2}s^{-1}}$, we obtain
$L_{k}\sim0.9\pm0.8\times10^{50}\,\mathrm{erg\, cm^{-2}s^{-1}}$.
Finally, for equation (\ref{eqnum3}) and $h\nu_{m,3}\sim
E_{pk}=1.0_{-0.4}^{+0.8}\,\mathrm{keV}$, we can obtain
$\gamma_{4}/\gamma_{1}^2\sim \mathrm{few}$, which is an essential
condition to produce a bright X-ray flare.

In addition, the optical depth due to pair production can be given
by \citep{lit01}
\begin{equation}
  {\tau _{\gamma \gamma }} = \frac{{(11/180){\sigma _T}{N_{ > {\nu _{m,an}}}}}}{{4\pi
  R^2}},
\end{equation}
where ${N_{ > {\nu _{m,an}}}}$ is the photon number with frequency
up to ${\nu _{m,an}}$ with $h\nu _{m,an}\equiv (\gamma m_e
c^2)^2/h{\nu _{m}}$, which can annihilate the ${\nu _{m}}\sim
1\,\mathrm{keV}$ photons. So ${N_{ > {\nu _{m,an}}}}\simeq L_{GeV} t
/(h{\nu _{m,an}})$ can be used to estimate the photon number with
frequency up to ${\nu _{m,an}}$, where $L_{GeV}$ is the GeV
luminosity. Besides, $R=2\gamma^2c\delta t\sim 2 \times
10^{14}\,\mathrm{cm}$, so we can get
  \begin{equation}
  {\tau _{\gamma \gamma }} \sim 2 \times {10^{ - 3}}L_{GeV,50}\gamma_{,1}^{ - 6}\delta {t^{-2}_{,2}}{t_{,
  1}},
  \end{equation}
which indicates that the pair production effect is unimportant.
As a result, the secondary electrons produced by the pair production
effect is ignored here.

To summarize, the GeV  emission of GRB100728A can be described well
by the IC process of the electrons accelerated by forward-reverse
shocks in regions 2 and 3. Using reasonable and appropriate values
of the model parameters, we present good fitting results (Fig.
\ref{f4}).

\subsection{Numerical Calculations of the Model}
The results mentioned above are analytical estimates, while all the
figures except for Figures 1 and 2 in this paper are based on more
detailed and precise numerical calculations. Next we will describe
numerical methods.

As mentioned above, the electrons accelerated by the shocks are
assumed to have a power-law energy distribution, $d{n'_{e}}/d{\gamma
'_{e}} \propto {\gamma '_e}^{ - p}$ for ${\gamma '_{e}} \ge {\gamma
'_{e,m}}$, where ${\gamma '_{e,m}}$ is the minimum Lorentz factor.
When the electron cooling effect is considered, the resulting
electron distribution in the comoving frame takes the following
forms:
\newline(1) If the newly shocked electrons cool faster than the shock
dynamical timescale, i.e. fast cooling ($\gamma '_{e,m} >
\gamma'_{e,c}$),
\begin{equation}
\frac{{d{n '_e}}}{{d{\gamma '_e}}} \propto \left\{ \begin{array}{ll}
{\gamma '_e}^{ - 2}  &   {\gamma '_{e,c}} < {\gamma '_e} < {\gamma '_{e,m}}\\
{\gamma '_e}^{ - p - 1}  &   {\gamma '_{e,m}} < {\gamma '_e} < {\gamma '_{e,\max }}  ;
\end{array} \right.
\end{equation}
\newline(2) If the newly shocked electrons cool slower than the shock
dynamical timescale, i.e. slow cooling ($\gamma '_{e,m} < \gamma
'_{e,c}$),
\begin{equation}
\frac{{d{n '_e}}}{{d{\gamma '_e}}} \propto \left\{ \begin{array}{ll}
{\gamma '_e}^{ - p}  &   {\gamma '_{e,m}} < {\gamma '_e} < {\gamma '_{e,c}}\\
{\gamma '_e}^{ - p - 1}  &   {\gamma '_{e,c}} < {\gamma '_e} < {\gamma '_{e,\max }}  ,
\end{array} \right.
\end{equation}
where ${\gamma '_{e,\max }}$ is the maximum Lorentz factor of
shocked electrons in the comoving frame, which is determined by
equating the electron acceleration timescale with the timescale of
the non-thermal emission (including synchrotron and IC emission)
cooling timescale.

From the electron distribution, the synchrotron seed photon spectrum
can be obtained easily \citep{rybicki79}. After we obtain the
electron distribution and the seed photon spectrum, the emission of
seed synchrotron photons up-scattered by relativistic electrons
accelerated by forward-reverse shocks can be computed. For
simplicity, we only consider the first-order IC and neglect higher
order IC processes. In the Thomson regime, therefore, the IC volume
emissivity in the comoving frame can be given by \citep{rybicki79,
sar01}
\begin{equation}
{j'}_\nu ^{IC} = 3{\sigma _T}\int_{{\gamma '_{e,m}}}^{\gamma '_{e,max}}
{d{\gamma '_e}} \frac{{d{n '_e}}}{{d{\gamma '_e}}}\int_0^1 {dxg(x){{\tilde f'}_{{{\nu '}_s}}}(x)} ,  \label{eqjnu}
\end{equation}
where $x=\nu'/(4 {{\gamma '}_e^2} \nu '_s)$ , $\nu '_s$ is the
synchrotron seed photon frequency in the comoving frame, ${{\tilde
f'}_{{{\nu '}_s}}}(x)$ is the incident-specific flux at the shock
front in the comoving frame, and $g(x)=1+x+2x\ln x-2 x^2$ considers
the angular dependence of the scattering cross section in the limit
$\gamma '_e \gg 1$ \citep{blumenthal70, sar01}. We can convert the
comoving-frame quantities to observed quantities, by considering
${f'}_\nu ^{IC} = {j'}_\nu ^{IC}4\pi {R^2}\Delta R'/4\pi {D^2}$ and
${{f'}_{\nu '}} = {{\tilde f'}_{{{\nu '}_s}}}4\pi {R^2}/4\pi {D^2}$,
where $R$ is the shock radius, $D$ is the distance to the observer,
$\Delta R'$ is the comoving width of the shocked shell \citep{sar01,
wang01b}. So we obtain the IC flux in the observer frame,
\begin{equation}
f_\nu ^{IC} = 3\Delta R'{\sigma _T}\int_{{{\gamma '}_{e,m}}}^{{{\gamma '}_{e,\max }}}
{d{{\gamma '}_e}} \frac{{d{{n'}_e}}}{{d{{\gamma '}_e}}}\int_0^1 {dxg(x){f_\nu }(x)}.
\end{equation}

If ${\gamma '_e}h{\nu '_s} \gtrsim {m_e}c^2$, the Klein-Nishina
regime should be considered. Equation (\ref{eqjnu}) can be replaced
by \citep{blumenthal70}
\begin{equation}
{j'}_\nu ^{IC} = 3{\sigma _T}\int_{{{\gamma '}_{e,m}}}^{{{\gamma '}_{e,\max }}}
{d{{\gamma '}_e}} \frac{{d{{n'}_e}}}{{d{{\gamma '}_e}}}\int_{{{\nu '}_{s,\min }}}^\infty
{d{{\nu '}_s}\frac{{\nu '{{\tilde f'}_{{{\nu '}_s}}}}}{{4{\gamma ^2}{\nu '}_s^2}}
[1 + x + 2x\ln x - 2{x^2} + \frac{1}{2}\frac{{{x^2}{y^2}}}{{1 + xy}}(1 - x)]} ,
\end{equation}
where $y=4{\gamma '_e} h \nu'_s / (m_e c^2)$ and $x=h \nu'/[y({\gamma '_e}m_e c^2
-h \nu' )]=\nu'/(4{{\gamma '}_e^2}\nu'_s -y\nu')$.

\subsection{Light Curves of the Model}
We now calculate synchrotron and IC emission light curves. It can be
predicted that both emissions will have a good temporal coincidence,
because they are produced from the same region. This may be the most
important difference from the EIC model, because in the latter model
the GeV emission will last for a period much longer than the
duration of the GeV emission based on  the curvature effect of an
external forward shock and is mainly extended by the highly
anisotropic radiation of the upscattered photons.

\citet{yu09} presented the theoretical X-ray flare light curves
produced by considering a collision of two homogeneous shells. Here
we give both X-ray and GeV emission light curves based on more
precise numerical calculations in our assumed dynamics in Fig.
\ref{f56}. A basic characteristic of the X-ray flare is that its
light curve has a rapid rise and fall. The rapid rise can be clearly
seen by resetting the time zero point in the right panel of Fig.
\ref{f56}. Before the two shocks' crossing time $T_{crs}$, by
ignoring possible spreading of the hot shocked materials, evolutions
of $\nu_{c,i}$, $\nu_{m,i}$, and $F_{\nu,max,i}$ follow equations
(\ref{eqnum2}), (\ref{eqnum3}), (\ref{eqnuc}), (\ref{fnumax2}), and
(\ref{fnumax3}). After $T_{crs}$, the spreading of the hot materials
into the vacuum cannot be ignored and the merged shell experiences
an adiabatic cooling. During this phase, a simple power-law of the
volume of the merged shell is assumed as $V '_i \propto R ^s $,
where $s$ is a free parameter and its value is taken to be from 2 to
3. As a result, the particle number densities would decrease as $n
'_i \propto {V '}_i^{-1} \propto R ^{-s}$, the internal energy
densities as $e '_i \propto {V '}_i^{-4/3} \propto R ^{-4s/3} $, and
the magnetic field strength as $B'_i \propto (e '_i)^{-1/2} \propto
R ^{-2s/3} $. From equation (\ref{R}), before $\delta t$, any
increase of the radius $R$ can be ignored (i.e., $R \simeq {\rm
constant}$), but after $\delta t$, the radius increases linearly
with time (i.e., $R \propto t$). For simplicity, we consider
$T_{crs}\simeq \delta t$. So the characteristic quantities can be
presented by
\begin{equation}
{\nu _m} \propto \left\{ \begin{array}{ll}
{t^0}    &    t < {T_{crs}}\\
{t^{ - 2s/3}}    &    t > {T_{crs}}, \label{numt}
\end{array} \right.
\end{equation}
\begin{equation}
{\nu _c} \propto \left\{ \begin{array}{ll}
{t^{ - 2}}    &     t < {T_{crs}}\\
{t^{2s - 2}}    &    t > {T_{crs}},   \label{nuct}
\end{array} \right.
\end{equation}
and
\begin{equation}
{F_{\nu ,\max }} \propto \left\{ \begin{array}{ll}
t     &     t < {T_{crs}}\\
{t^{ - 2s/3}}   &     t > {T_{crs}}.
\end{array} \right.    \label{fnummaxt}
\end{equation}
For clarity, the subscript $i$ is omitted. The theoretical light
curve of an X-ray flare has been given in appendix A.

The intrinsic decline slope of the last segment of the theoretical
light curves is $\alpha =(sp+s)/3$ (where $\alpha = -d \log F_\nu /d
\log t$). \citet{liang06} found that the rapid decline of most X-ray
flares seems to be consistent with the curvature effect by fitting
the light curves of X-ray flares detected by Swift and that the
temporal index is equal to the simultaneous spectral index plus $2$.
In the last segment of the theoretical light curves, the
corresponding spectral index is $(p-1)/2$ for
$\nu_m<\nu_{X_{band}}<\nu_c$, where $\nu_{X_{band}}\sim10^{17}
\mathrm{Hz}$. For $s=3$ and $2<p<3$, we find
\begin{equation}
\alpha  = \frac{{3p + 3}}{3} > \frac{{p - 1}}{2} + 2.  \label{alpha}
\end{equation}
So the X-ray flux would have a rapid decline owing to the curvature
effect.

Similarly, in the left panel of Fig. \ref{f56}, several apparent
power-law forms are written as
\begin{equation}
{F_\nu } \propto \left\{ \begin{array}{ll}
t     &    t < {T_{cm}}\\
{t^0}    &    {T_{cm}} < t < {T_{crs}}\\
{t^{\frac{{s - 3}}{3}}}   &    {T_{crs}} < t < {T_m}\\
{t^{ - \frac{{sp - 2s + 3}}{3}}}    &    t > {T_m}.
\end{array} \right.
\end{equation}
The temporal index $\alpha$ of the last segment of the light curves
is $(sp-2s+3)/3$. Although this temporal index cannot easily satisfy
equation (\ref{alpha}), it cannot be ruled out absolutely. This is
because the segment near the flare onset time may be steepened by
the time zero effect dramatically. This effect can be seen by
comparing the right panel with the left panel of Fig. \ref{f56},
where the right panel resets the time zero point, having a larger
slope. So X-ray flares formed by two-shell interactions are
characteristic of a rapid rise and fall.

In Fig. \ref{f56}, the X-ray and GeV emission have a similar
evolution with time, which can be easily seen in the right panel. It
is this behavior that we want to specify both time coincidence.

\section{Conclusions}

In this paper, the late internal shock origin for X-ray flares is
adopted, and a collision of two homogeneous shells is analyzed in
quantitative calculations. Besides this model, X-ray flares may be
produced by a delayed external shock \citep{piro05, galli07}. Both
models suggest a prolonged central engine activity. \citet{wu05}
made a quantitative analysis in two cases, and suggested that two
kinds of X-ray flares are not excluded, and even maybe coexist for a
certain GRB.

The strong SSC and CIC emission during X-ray flares had been
analyzed and found to be detectable with high energy telescopes
\citep{fan08,yu09}. GRB 100728A is the second case (after GRB090510)
to date with simultaneous Swift and Fermi observations, in which the
GeV and X-ray emission maybe have the same origin because of the
temporal coincidence. Thus, the afterglow synchrotron and SSC
emission scenarios may be slightly far-fetched. It is natural that
high energy emission can be generated during X-ray flares by inverse
Compton processes. \citet{he12} provided an explanation for GRB
100728A in the EIC scenario, in which X-ray flare photons are
up-scattered by electrons in an external forward shock. We here give
an alternative reasonable explanation by using the SSC and CIC
scenario where X-ray flare photons are up-scattered by electrons
accelerated by forward-reverse shocks in the late internal shock
model. One main difference between the two scenarios is whether
there is a good temporal correlation between X-ray and GeV emission
\citep{fan08}. In the SSC and CIC scenario, a good temporal
correlation between X-ray and GeV emission is expected (Fig.
\ref{f56}), whereas GeV photons in the EIC scenario maybe have a
significant temporal extension and even last a time much longer than
the duration of one X-ray flare \citep{fan08}. So, no obvious
temporal extension of GeV photons for GRB 100728A supports the SSC
and CIC scenario. In fact, both the SSC and CIC scenario and the EIC
scenario are not excluded and maybe coexist in high energy emission,
because the extended GeV emission flux in the EIC scenario may be
too weak (compared with that in the SSC and CIC scenario) to be
detected.

\acknowledgments We thank the referee for helpful comments and
constructive suggestions that have allowed us to improve the
manuscript significantly, and Yunwei Yu for useful discussions. This
work was supported by the National Natural Science Foundation of
China (grant no. 11033002).

\clearpage

\appendix

\section{Appendix}

\begin{figure}
\epsscale{.96} \plotone{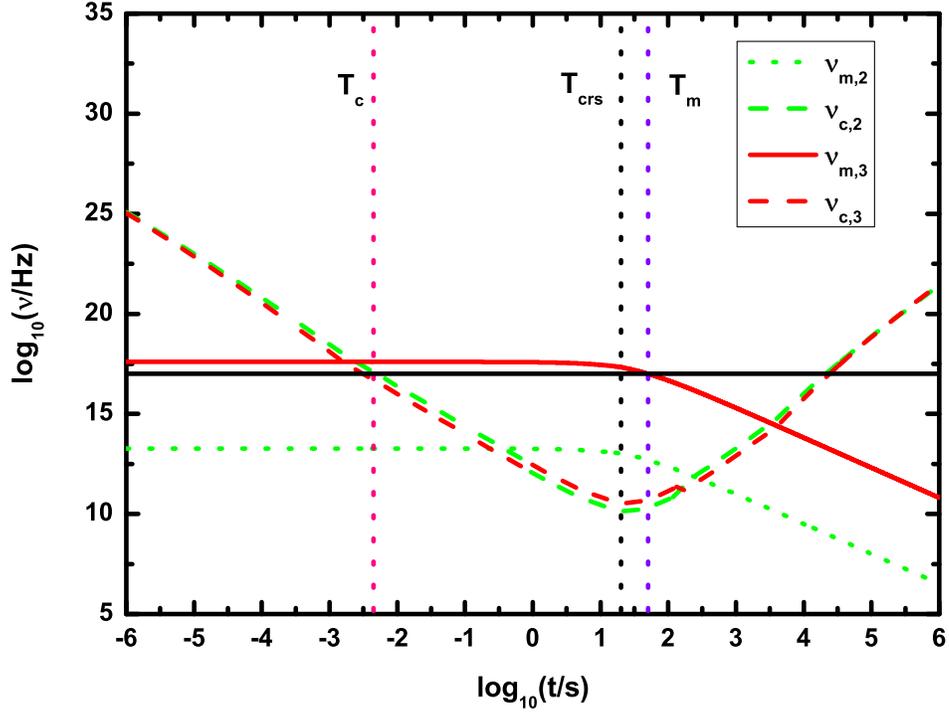} \caption{Four characteristic
frequencies as functions of time. The black vertical dotted line
represents the forward and reverse shock crossing time. A similar
crossing time $T_{crs}=20\,\mathrm{s}$ of two shocks is also
assumed. After the shock crossing time, the merged shell expands
adiabatically if $s=3$ is assumed. $T_m$($T_c$) is the time of the
break frequency $\nu_m$($\nu_c$) passing through the X-ray band
(black horizontal solid line, $~10^{17}\,\mathrm{Hz}$) in the region
3 \citep{yu09}. The same parameters as in Fig. \ref{f1} are taken in
numerical calculations.\label{f2}}
\end{figure}

\begin{figure}
\epsscale{.96} \plotone{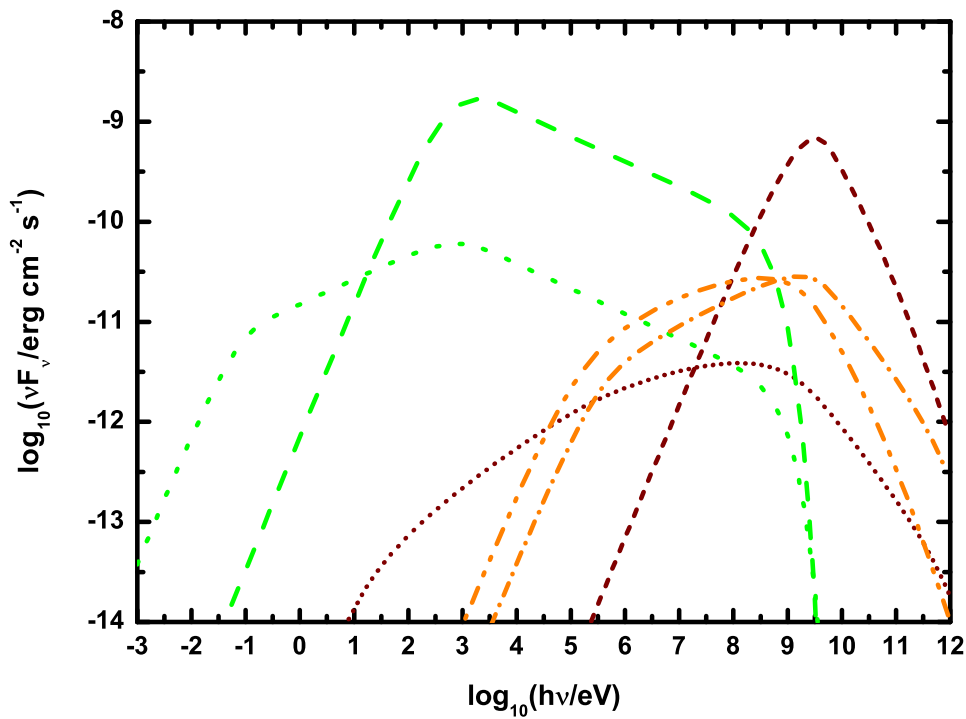} \caption{Time-resolved spectra of
the six components at $t=10$\,ms. The green dotted line and dashed
line represent the synchrotron emission of regions 2 and 3,
respectively. The wine short-dotted line and short-dashed line
represent the SSC emission of regions 2 and 3, respectively. The
orange dash-dot-dotted line and short-dash-dotted line represent the
CIC emission, respectively. The same parameters as in Fig. \ref{f1}
are taken in numerical calculations.\label{f3}}
\end{figure}

\begin{figure}
\epsscale{.96} \plotone{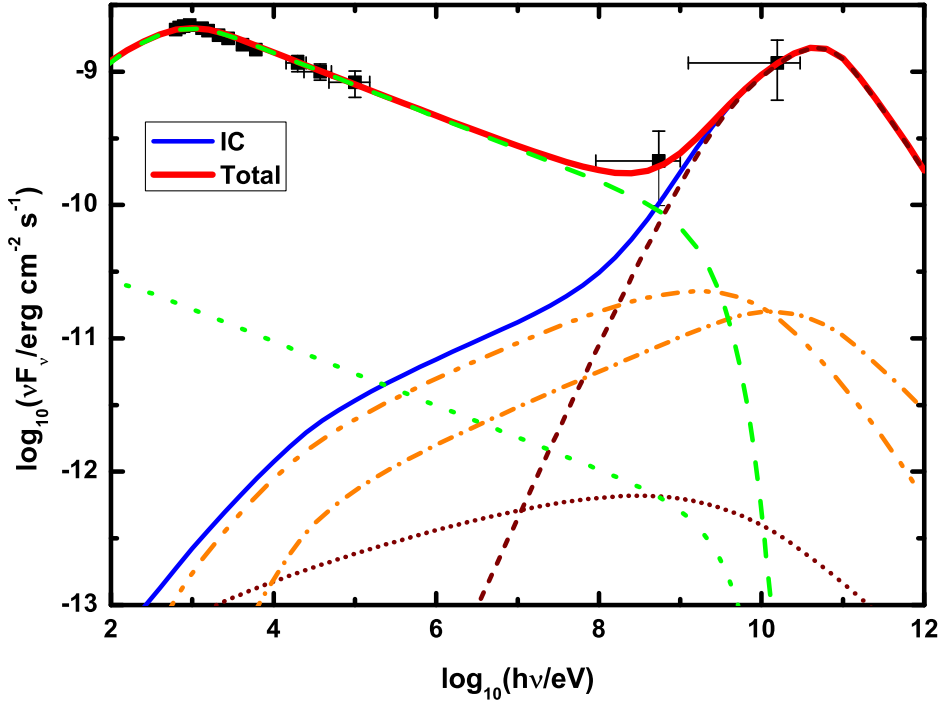} \caption{Time-averaged spectra of
the X-ray and GeV emission of GRB 100728A and fitting this burst
with our model. The observed data are taken from \citet{abd11},
which are fitted by a time-averaged spectrum from $t=0\,\mathrm{s}$
to $t=10^{1.8}\,\mathrm{s}$. The green dotted line and dashed line
represent the synchrotron emission of regions 2 and 3, respectively.
The wine short-dotted line and short-dashed line represent the SSC
emission of regions 2 and 3, respectively. The orange
dash-dot-dotted line and short-dash-dotted line represent the CIC
emission, respectively. The blue thin solid line represents the
total IC including SSC and CIC and the red thick solid line
represents the sum of synchrotron and IC emission. The other
parameters $L_{k,1}=7.0\times10^{50}\mathrm{erg s^{-1}}$,
$L_{k,4}=2.5\times10^{50}\mathrm{erg s^{-1}}$, $\gamma_{1}=50$,
$\gamma_{4}=5830$, $p=2.48$, $\epsilon_e=0.3$, $\epsilon_B=0.03$,
$\theta_{jet}=0.1$, and $z=1$ are taken in numerical calculations.
\label{f4}}
\end{figure}

\begin{figure}
\epsscale{.96} \plottwo{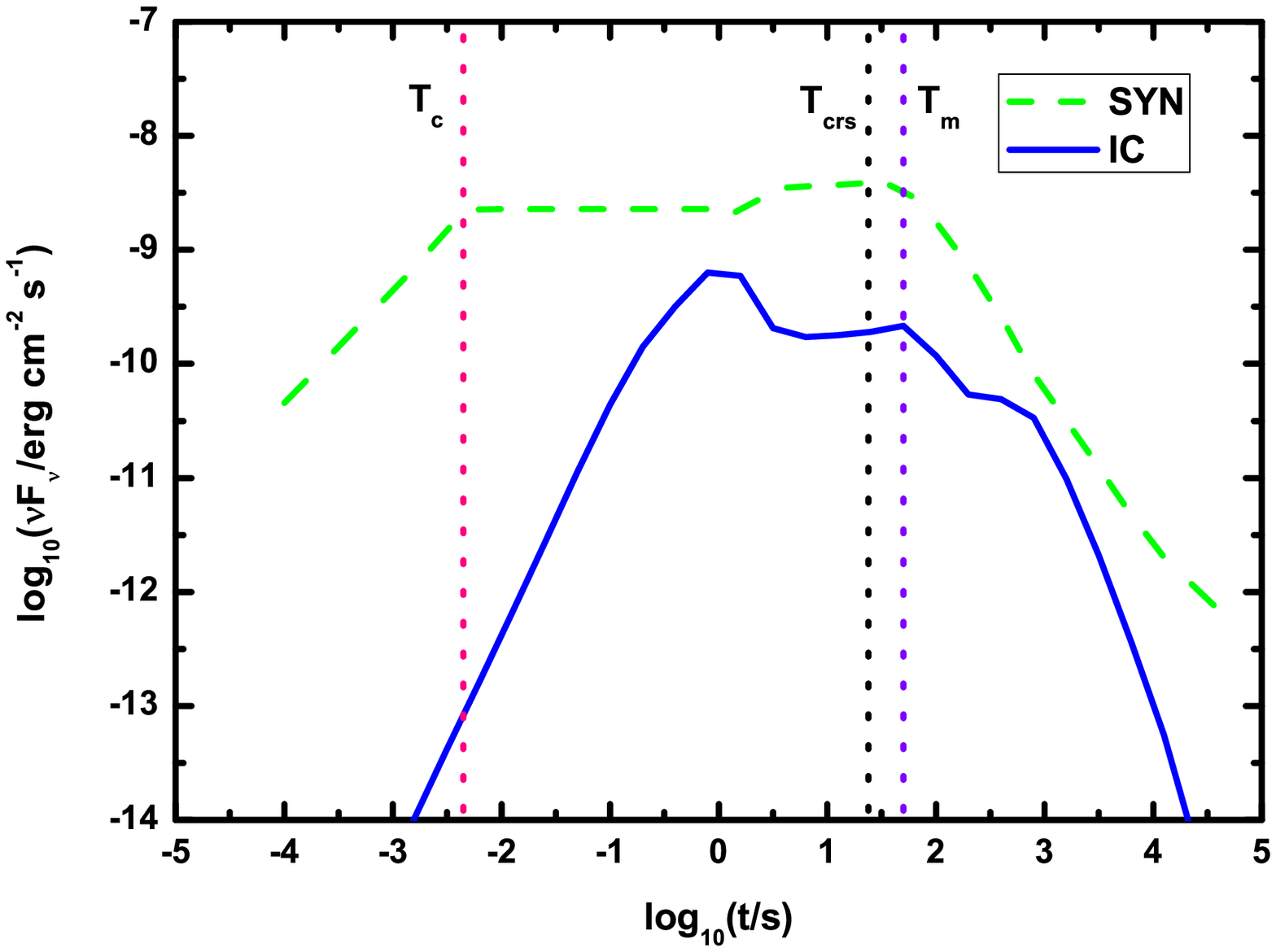}{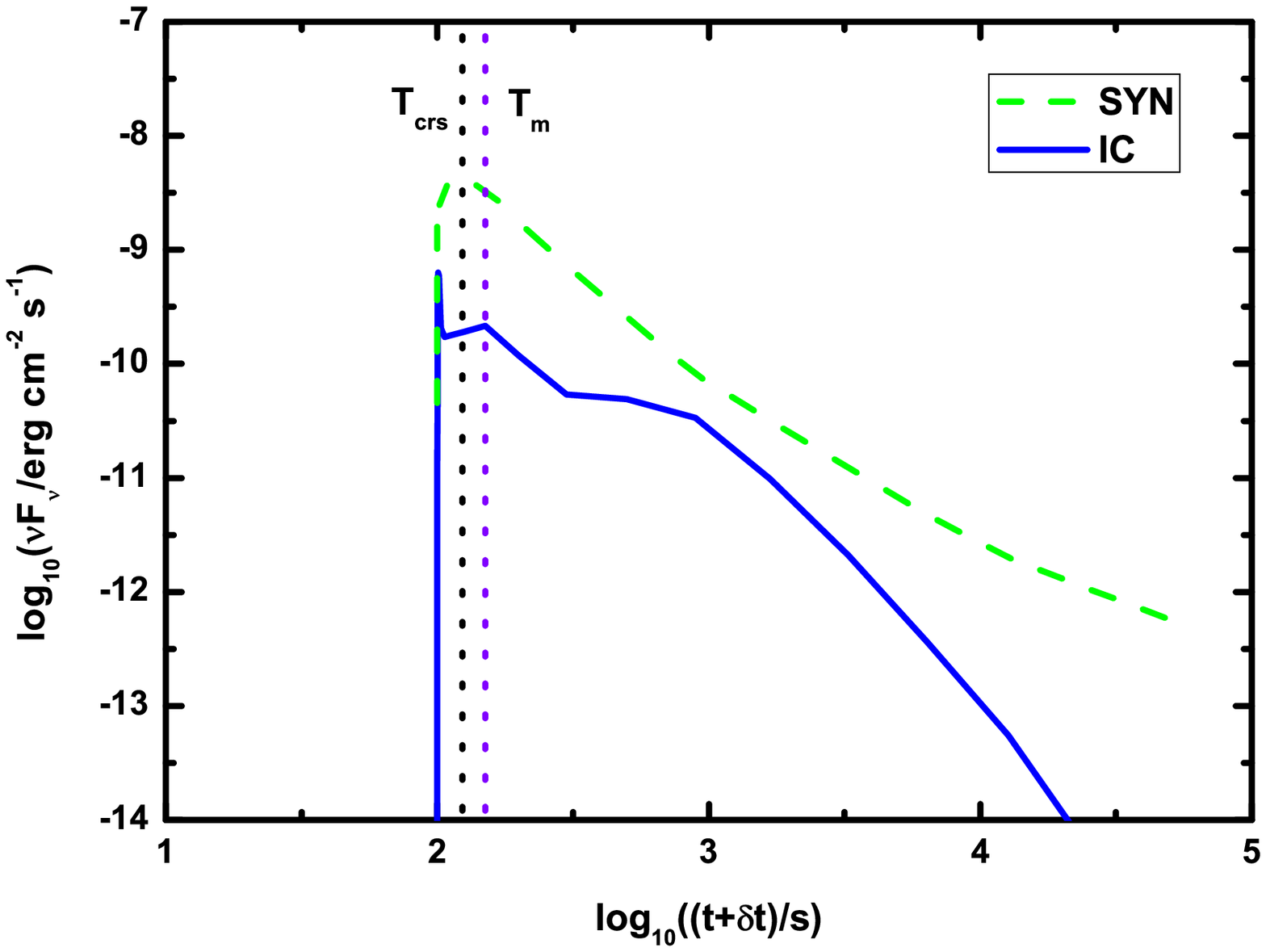} \caption{The light lines
of the synchrotron and IC emission. The left panel is the flux verse
time, and the right panel is obtained by resetting the time zero,
where $\delta t=100\,$s is assumed. The green dashed line and the
blue solid line represent the synchrotron emission and IC emission
respectively. The three vertical dotted lines have the same meaning
as in Fig. \ref{f2}. The same parameters as in Fig. \ref{f1} are
taken in numerical calculations.\label{f56}}
\end{figure}

Here we present the theoretical X-ray flare light curves in the
parameters of Fig. \ref{f2}. The synchrotron energy spectrum can
obtained from \citet{sar98}.
\newline(1) In the fast cooling regime, the energy spectrum is discribed by
\begin{equation}
{F_\nu } = \left\{ \begin{array}{ll}
{(\nu /{\nu _c})^{1/3}}{F_{\nu ,\max }}   &    \nu  < {\nu _c}\\
{(\nu /{\nu _{c}})^{ - 1/2}}{F_{\nu ,\max }}      &   {\nu _c} < \nu  < {\nu _m}\\
{(\nu /{\nu _m})^{ - p/2}}{({\nu _m}/{\nu _c})^{ - 1/2}}{F_{\nu ,\max }}    &   {\nu _m} < \nu  ;
\end{array} \right. \label{fastcooling}
\end{equation}
\newline(2) And for slow cooling, the energy spectrum reads
\begin{equation}
{F_\nu } = \left\{ \begin{array}{ll}
{(\nu /{\nu _m})^{1/3}}{F_{\nu ,\max }}    &   \nu  < {\nu _m}\\
{(\nu /{\nu _m})^{ - (p - 1)/2}}{F_{\nu ,\max }}     &     {\nu _m} < \nu  < {\nu _c}\\
{(\nu /{\nu _c})^{ - p/2}}{({\nu _c}/{\nu _m})^{ - (p - 1)/2}}{F_{\nu ,\max }}   &    {\nu _c} < \nu.
\end{array} \right.  \label{slowcooling}
\end{equation}
For a specific x-ray band in Fig. \ref{f2}, from equations and
equations (\ref{numt}), (\ref{nuct}), and (\ref{fnummaxt}), the
theoretical X-ray flare light curves can be given by
\begin{equation}
{F_\nu } \propto \left\{ \begin{array}{ll}
t          & t < {T_{cm1}}\\
{t^{5/3}} &  {T_{cm1}} < t < {T_{c1}}\\
{t^0}     &{T_{c1}} < t < {T_{crs}}\\
{t^{\frac{{s - 3}}{3}}}    &   {T_{crs}} < t < {T_m}\\
{t^{ - \frac{{sp - 2s + 3}}{3}}}   &    {T_m} < t < {T_{cm2}}\\
{t^{ - \frac{{sp - 2s + 3}}{3}}}    &    {T_{cm2}} < t < {T_{c2}}\\
{t^{ - \frac{{sp + s}}{3}}}    &     t > {T_{c2}}.
\end{array} \right.
\end{equation}
where $T_{cm1}(T_{cm2})$ is the first (second) time $\nu_c=\nu_m$,
and $T_m$($T_{c1}$ or $T_{c2}$, $T_{c1}$ for the first time and $T_{c2}$ for
the second time) is the time of the break frequency
$\nu_m$($\nu_{c}$) passing through the X-ray band (about $10^{17}\,
\mathrm{Hz}$) in region 3. It should be pointed out that there may
be a mistake in \citet{yu09}, which gave a temporal index $(sp+3)/3$
for $t>{T_{c2}}$.

\clearpage

\clearpage

\clearpage

\end{document}